\documentclass[10pt]{article}
\usepackage{graphicx}

\def\Title#1{\begin{center} {\Large #1 } \end{center}}
\def\Author#1{\begin{center}{ \sc #1} \end{center}}
\def\Address#1{\begin{center}{ \it #1} \end{center}}

\newcommand\pubblock{\rightline{\begin{tabular}{l} Proceedings of the Second Annual LHCP\\ \pubnumber\\
         \pubdate  \end{tabular}}}

\newenvironment{Abstract}{\begin{quotation} \begin{center} 
             \large ABSTRACT \end{center}\bigskip 
      \begin{center}\begin{large}}{\end{large}\end{center} \end{quotation}}

\newenvironment{Presented}{\begin{quotation} \begin{center} 
             PRESENTED AT\end{center}\bigskip 
      \begin{center}\begin{large}}{\end{large}\end{center} \end{quotation}}

\def\beq{\begin{equation}}
\def\eeq#1{\label{#1}\end{equation}}
\def\eeqn{\end{equation}}

\def\beqa{\begin{eqnarray}}
\def\eeqa#1{\label{#1}\end{eqnarray}}
\def\eeqan{\end{eqnarray}}

\let\bar=\overbar

\def\Dslash{\not{\hbox{\kern-4pt $D$}}}
\def\dslash{\not{\hbox{\kern-2pt $\del$}}}

\def\msb{{\bar{\ssstyle M \kern -1pt S}}}

\usepackage{color}

\newcommand\pubnumber{ ATL-PHYS-PROC-2014-133 }

\newcommand\pubdate{\today}

\def\affiliation{
On behalf of the ATLAS Experiment, \\
Department of Physics \\
University of Pennsylvania, Philadelphia, PA 19104, U.S.A }

\usepackage{hyperref}
\usepackage{xspace}
\usepackage{subfigure}
\usepackage{lineno}
\def\MeV{\ifmmode {\mathrm{\ Me\kern -0.1em V}}\else
                   \textrm{\,Me\kern -0.1em V}\xspace\fi}%
\def\GeV{\ifmmode {\mathrm{\ Ge\kern -0.1em V}}\else
                   \textrm{\,Ge\kern -0.1em V}\xspace\fi}%
\def\TeV{\ifmmode {\mathrm{\ Te\kern -0.1em V}}\else
                   \textrm{\,Te\kern -0.1em V}\xspace\fi}%
\def\ps{\ifmmode {\mathrm{\ ps}}\else
                  \textrm{\,ps}\xspace\fi}%
\def\ystar{\ensuremath{y^{\mathrm{*}}}\xspace}
\def\pt{\ensuremath{p_{\mathrm{T}}}\xspace}
\def\ptphi{\ensuremath{p_{\mathrm{T},\phi}}\xspace}

\begin{document}

\large
\begin{titlepage}
\pubblock

\vfill
\Title{  Recent QCD Results from ATLAS  }
\vfill

\Author{ Christopher J. Meyer }
\Address{\affiliation}
\vfill
\begin{Abstract}

Recent QCD results from ATLAS taken at 7\TeV center-of-mass energy using the LHC are presented, including: 
dijet production, isolated photon production, isolated photon production associated with jets, jet shapes in top-quark pair events, the production cross-section of the $\phi$(1020) meson, and underlying event in jet events.
Good agreement with theory predictions is seen, in particular with those made by next-to-leading-order generators.
These measurements highlight the importance of precision QCD measurements for improving state-of-the-art theoretical tools and searching for new physics.

\end{Abstract}
\vfill

\begin{Presented}
The Second Annual Conference\\
 on Large Hadron Collider Physics \\
Columbia University, New York, U.S.A \\ 
June 2-7, 2014
\end{Presented}
\vfill
\end{titlepage}
\def\thefootnote{\fnsymbol{footnote}}
\setcounter{footnote}{0}
%

\normalsize 


\section{Introduction}

Detailed checks of quantum chromodynamics (QCD) are a crucial part of current high energy particle physics research programs.
Precise measurements of jets, photons, and a combination of the two provide powerful constraints on the gluon content of the proton.
By including these measurements in combined global fits, the uncertainty due to the proton parton distribution function (PDF) at high momentum fraction can be significantly reduced.
This benefits not only other Standard Model (SM) measurements, including those of the Higgs boson, but also searches for new physics.
Similarly, most analyses are affected by the complex underlying event (UE) present in all proton--proton collisions.
A better understanding of the UE as well as the parton shower will lead to an improved description by MC generators such as PYTHIA and HERWIG, which are used in a wide range of studies.
Finally, these results can be used to confront next-to-leading-order (NLO) predictions, testing how well the SM describes regions of phase space never before measured.
Any significant discrepancy between measurements and SM predictions could be the first sign of new physics.

A brief description of dijet production \cite{Aad:2013tea}, prompt photon production (with \cite{Aad:2013gaa} and without \cite{Aad:2013zba} associated jets), jet shapes in top-quark pair events \cite{Aad:2013fba}, $\phi$(1020)-meson production \cite{Aad:2014rca}, and UE in jet events \cite{Aad:2014hia} from ATLAS \cite{Aad:2008zzm} at the LHC is presented in the following sections.
Finally, a conclusion on the various results is given.

\section{Dijet production}

Dijet production is important for testing NLO QCD predictions and constraining the gluon PDF, as well as confronting new physics.
The anti-$k_t$ clustering algorithm is used to define jets from three dimensional energy deposits in the ATLAS calorimeters.
Values of the radius parameter $R=0.4$ and $R=0.6$ are used to provide sensitivity to the parton shower and UE, respectively.
A data-driven calibration of the jets is performed, correcting both their energy and direction due to detector effects such as calorimeter non-compensation and dead material.
Events with at least two jets within $|y|<3.0$ are considered, with an additional requirement of $\pt > 100\ (50) \GeV$ on the highest (second highest) \pt jet.

\begin{figure}[htb]
\centering
\includegraphics[width=0.9\linewidth]{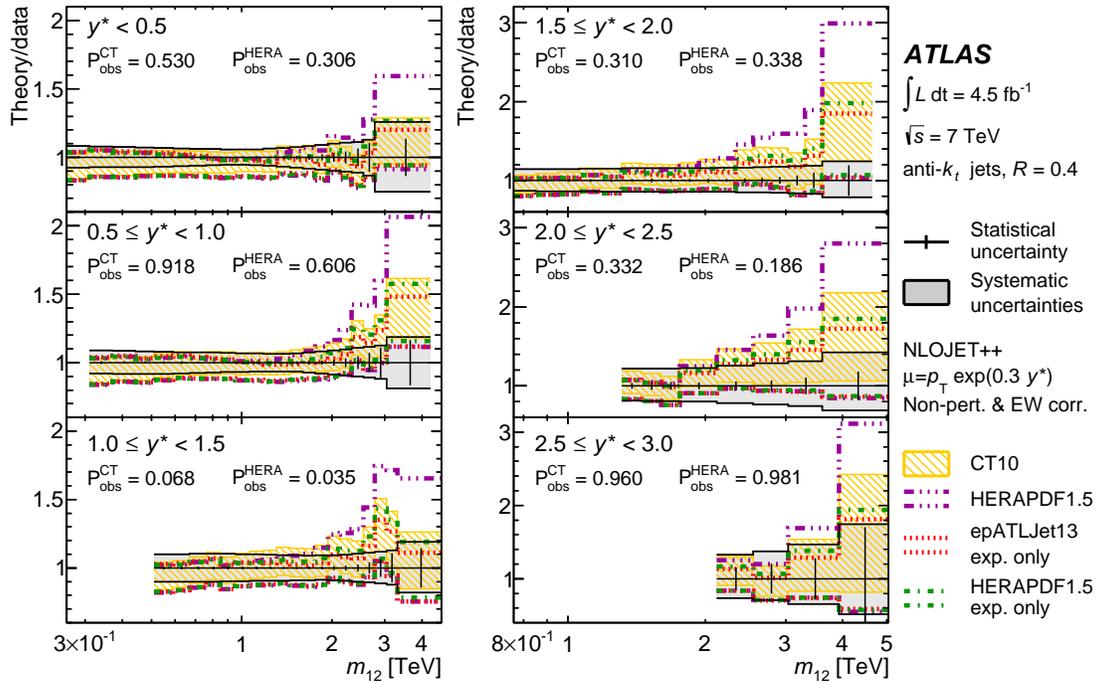}
\caption{
Comparison of measured and expected dijet production cross-sections, as a function of dijet mass in different ranges of \ystar \cite{Aad:2013tea}.
}
\label{fig:dijets}
\end{figure}

Dijet cross-sections are measured as functions of dijet mass and half the rapidity separation $\ystar = |y_1 - y_2|/2$ of the two highest-\pt jets.
All results are corrected back to the particle level, defined for particles with a proper lifetime greater than 10\ps.
The predictions of two NLO generators are compared with the data.
First, those of NLOJet++ are considered, which include non-perturbative corrections provided by the leading-order (LO) PYTHIA generator.
Second, those of POWHEG+PYTHIA, where an NLO matrix element is interfaced with a LO parton-shower generator.
In both cases, an additional correction for the effects of NLO electroweak diagrams is included.
Figure \ref{fig:dijets} shows a comparison of data with NLOJet++ predictions using the HERAPDF1.5 and CT10 PDF sets.

A quantitative comparison, using a Bayesian technique with an asymmetric $\chi^2$, shows good agreement between NLOJet++ predictions and data.
However, in a few ranges of \ystar the HERAPDF1.5 PDF set shows poor agreement with data, and the ABM11 PDF set fails to describe the data for values of $\ystar < 3.0$.
The $CLs$ technique is used on the publicly available results to confront a NLO model of contact interactions, excluding compositness scales $\Lambda < 6.9$--$7.7\TeV$, depending on the PDF set considered.

\section{Isolated photon production}

Photons are measured from electromagnetic deposits in the ATLAS calorimeters, and calibrated using a data-driven method based on the $Z$-boson mass peak.
Their selection is further refined using an identification based on their three-dimensional shower shape in the calorimeters.
Only prompt photons are considered, which includes those from direct production as well as fragmentation production.
This excludes all photons from hadronic decays, which are rejected by requiring minimal additional activity in the calorimeters surrounding prompt photon candidates.
All results are corrected to the particle level, accounting for detector resolution effects.

\begin{figure}[htb]
\centering
\subfigure[ ]{\label{fig:photons:a}\includegraphics[width=0.45\linewidth]{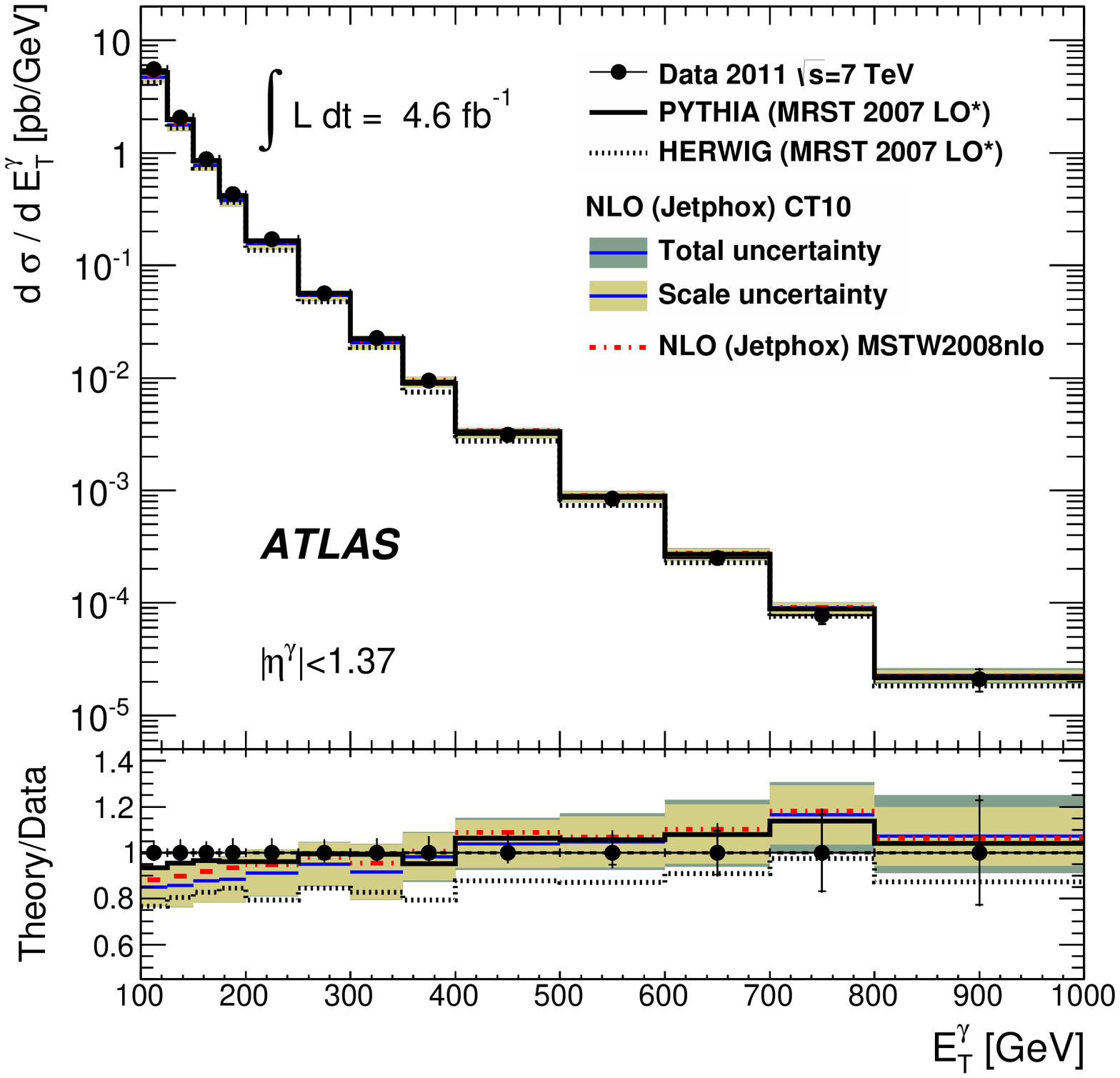}}
\subfigure[ ]{\label{fig:photons:b}\includegraphics[width=0.45\linewidth]{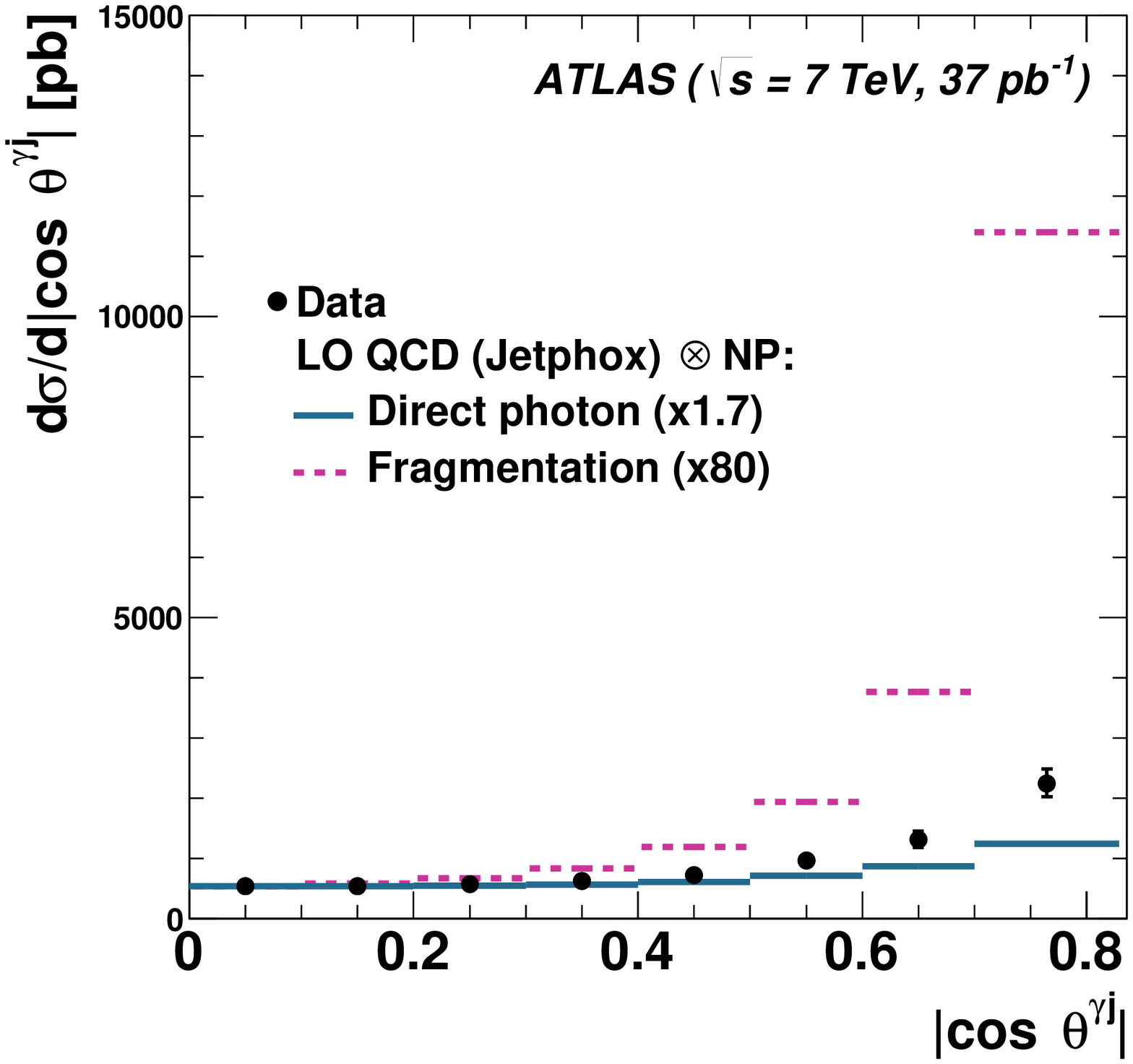}}
\caption{
In (a) the comparison of measured and expected prompt photon cross-sections, as a function of the transverse energy of photons in range $|\eta^\gamma|<1.37$ \cite{Aad:2013zba}.
In (b) the comparison of measured $|\cos{\theta^{\gamma\mathrm{j}}}|$ with LO predictions of direct and fragmentation photon production \cite{Aad:2013gaa}.
}
\label{fig:photons}
\end{figure}

For the inclusive prompt photon measurement, there are additional requirements of $100\GeV < E_\mathrm{T}^\gamma < 1000\GeV$, $|\eta| < 2.37$ (excluding the crack region $1.37 < |\eta| < 1.52$), and an isolation requirement $E_\mathrm{T,iso} < 7\GeV$ within $\Delta R < 0.4$ around the photon.
The non-prompt background is subtracted using the ``two-dimensional side bands'' method --- an extrapolation in isolation and particle identification --- extended to account for the migrations between signal and background samples.
Cross-sections are shown as a function of $E_\mathrm{T}^\gamma$ in Figure \ref{fig:photons:a} for the range $|\eta^\gamma| < 1.37$.
Theoretical predictions by Jetphox, a NLO calculation including both direct and fragmentation photon production, are shown using the CT10 and MSTW2008 PDF sets.
Good agreement is observed for higher values of transverse energies, although Jetphox underpredicts the data at lower values of transverse energy.
Since a similar, but exaggerated, trend is observed for LO predictions of the direct production component made by PYTHIA and HERWIG, this emphasizes the importance of higher-order fragmentation terms.
The primary production channel is through $u$-quark and gluon scattering.
Comparing the gluon PDF between different analyses \cite{ATL-PHYS-PUB-2013-018}, differences are seen across the entire range of parton momentum fraction, emphasizing the importance of this data to constrain future fits.

When studying the dynamics of events with photons produced in association with jets, at least one anti-$k_t$ jet with radius parameter $R=0.6$ is required.
Similar kinematic selections of photons and background subtraction are used as those in the inclusive prompt photon measurement.
The isolation requirement is decreased to $E_\mathrm{T,iso} < 3\GeV$ within a radius of $\Delta R < 0.4$ around the photon.
In addition to predictions of direct and fragmentation production at NLO by Jetphox, bin-to-bin corrections accounting for non-perturbative corrections are included.
Jetphox predictions describe both angular and scalar ($p_\mathrm{T}^\mathrm{jet}$, $E_\mathrm{T}^\gamma$) distributions well.
In Figure \ref{fig:photons:b} the angular distribution of $|\cos{\theta^{\gamma\mathrm{j}}}|$ is shown, compared with LO Jetphox predictions of the direct and fragmentation contributions separately.
This again emphasizes the importance of including both photon production modes in the theory calculation.

\section{Jet shapes in top-quark pair events}

Due to differences between $ggg$ and $qqg$ vertices, gluons lead to more parton radiation, making gluon initiated jets broader than quark initiated jets.
Jets initiated by $b$-quarks ($b$-jets) are expected to be broader than those initiated by $u$- or $d$-quarks (light jets) due to their larger mass.
Because $t\rightarrow W b$ almost exclusively, examining top-quark pair events provides an enriched $b$-jet sample.
Both a two and one lepton sample are considered, depending on whether both W bosons decay as $W\rightarrow l \nu$, or whether one decays hadronically to two light jets.
Here, the lepton can be either an electron or a muon, and missing transverse energy is used as a proxy to the neutrino.
The single lepton channel provides an enriched sample of light jets which can be compared with the $b$ jets.

\begin{figure}[htb]
\centering
\includegraphics[width=0.75\linewidth]{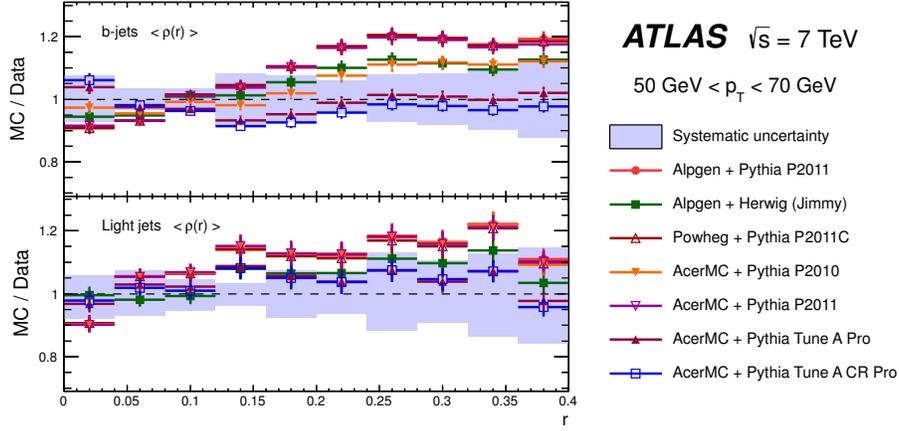}
\caption{
Comparison of measured and expected differential jet shape, for jets with $50\GeV < \pt < 70\GeV$ \cite{Aad:2013fba}.
}
\label{fig:jetshapes}
\end{figure}

Jets are defined using the anti-$k_t$ algorithm, with radius parameter $R=0.4$, and are corrected in the same way as those used to measure dijet production.
The shape of jets is defined as the \pt present in annuli of width $\Delta R = 0.04$ around the jet, a distance $r$ (in $\eta/\phi$-space) from the jet axis.
Each value is normalized by the width of the annuli and the total \pt of the jet, then averaged across the entire sample of jets.
A comparison of different MC generators with data is shown in Figure \ref{fig:jetshapes}, where AcerMC+PYTHIA Tune A (CR) Pro is seen to provide the best description.
The predictions of MC@NLO+HERWIG and POWHEG+PYTHIA AMBT1 (not shown) are also found to describe the data well.

\section{Production cross-sections of the \texorpdfstring{$\phi$}{phi}(1020) meson}

The production cross-section of the $\phi$(1020) meson decaying to opposite sign Kaons $\phi \rightarrow K^+ K^-$ is measured using data collected in early 2010 running, where pileup conditions are negligible.
Kaons are identified through a likelihood method based on the energy loss of charged-particle tracks in the pixel detector.
In order to maintain a reasonable particle identification efficiency for the Kaons, the measurement is made in the fiducial region:
$p_\mathrm{T,K} > 230\MeV$,
$p_\mathrm{K} < 800\MeV$,
$500\MeV < p_{\mathrm{T,}\phi} < 1200\MeV$, and
$|y_\phi| < 0.8$.
The yield in each bin of the measurement is extracted by a signal$+$background fit of the di-Kaon mass distribution, and is corrected for detector inefficiency.

\begin{figure}[htb]
\centering
\includegraphics[width=0.71\linewidth]{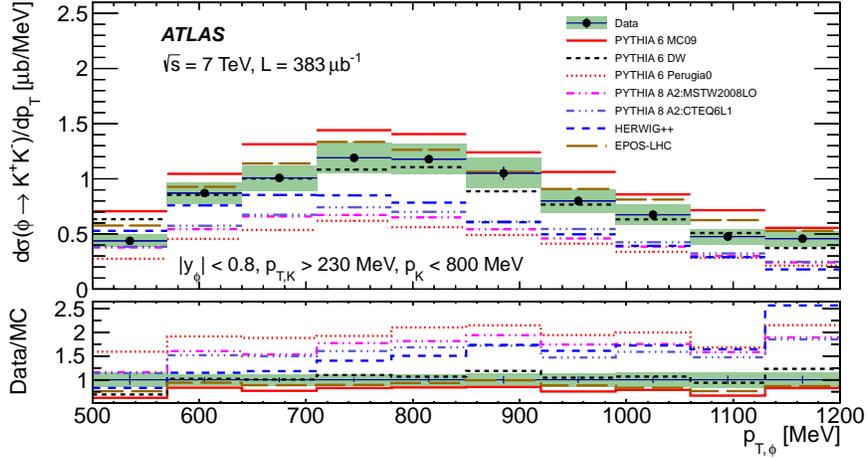}
\caption{
The $\phi$(1020)$\rightarrow K^+K^-$ fiducial cross section, as a function of \ptphi \cite{Aad:2014rca}.
}
\label{fig:phimeson}
\end{figure}

Production cross-sections are shown in Figure \ref{fig:phimeson} as a function of $p_{\mathrm{T,}\phi}$, along with predictions by various tunes of PYTHIA, as well as HERWIG++ and EPOS-LHC.
Both EPOS-LHC and PYTHIA6 DW tunes describe the data well, while other tunes either overestimate or underestimate the cross-section.

\section{Underlying event in jet events}

\begin{figure}[htb]
\centering
\subfigure[ ]{\label{fig:ujets:a}\includegraphics[height=0.25\paperheight]{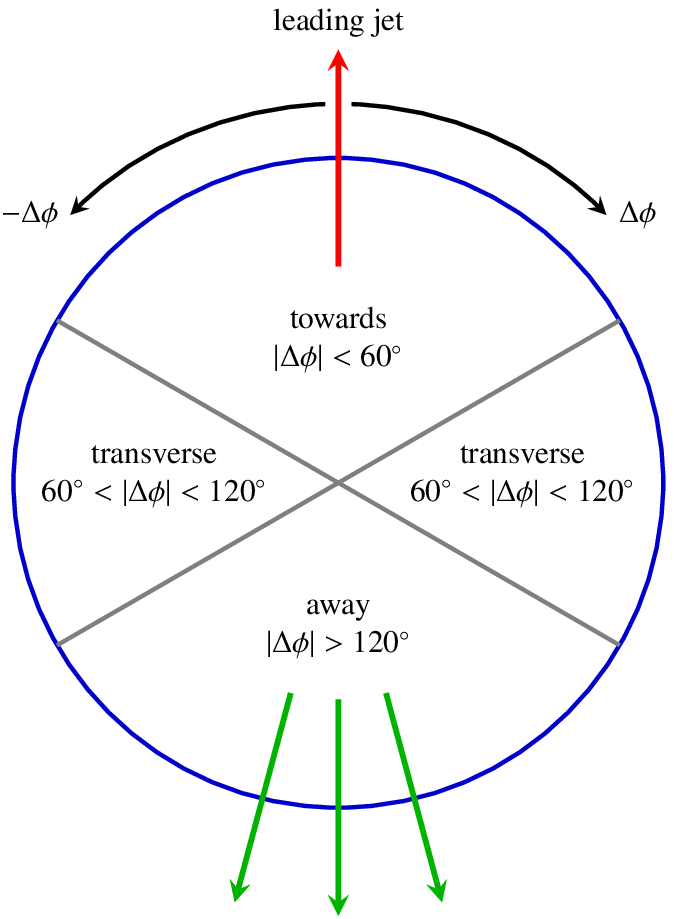}} \hspace{0.05\linewidth}
\subfigure[ ]{\label{fig:ujets:b}\includegraphics[height=0.25\paperheight]{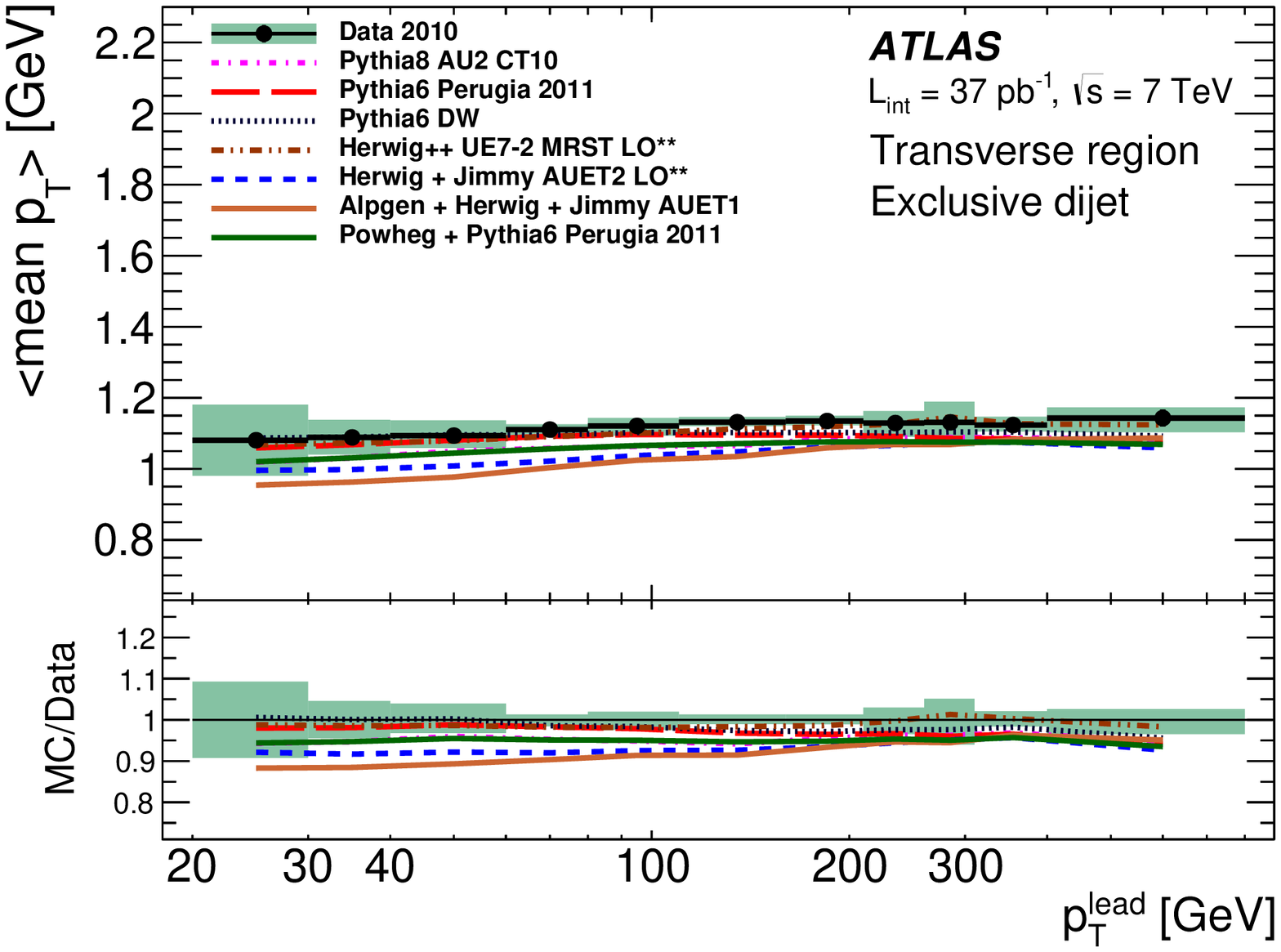}}
\caption{
In (a) the definition of the towards, away, and transverse regions is shown with respect to the leading jet direction.
In (b) measured and expected mean \pt in the transverse region are shown, as a function of leading jet \pt \cite{Aad:2014hia}.
}
\label{fig:uejets}
\end{figure}

When searching for new physics, it is important to understand the UE as well as the hard-scatter process.
Here, the UE includes multiple parton interactions, QCD color connections between partons and beam remnants, as well as initial and final state radiation.
Because perturbative QCD doesn't describe this low-momentum region, tuneable phenomenological models are necessary.
Events containing at least one calibrated, anti-$k_t$ jet with radius parameter $R=0.4$ are considered.
The event is split into three general $\phi$ regions, defined in Figure \ref{fig:ujets:a}: towards, away from, and transverse to the highest \pt jet in the event.

The average of the distribution of mean charged-particle \pt is studied among other observables, all corrected for detector effects.
Figure \ref{fig:ujets:b} shows this mean as a function of leading jet \pt, which is flat in the transverse region when restricting to events with exactly two jets.
While this points towards UE reaching a plateau for higher energy collisions, an expected result, other distributions observe a slight decrease at high \pt, warranting further studying.
The PYTHIA and HERWIG MC generators are seen to give a good basic description of the data, with different tunes working better or worse for different observables.
In particular, PYTHIA8 is seen to predict an UE with too high-\pt charged particles, implying an improved tune will benefit future versions.

\section{Conclusions}

In general, NLO QCD calculations are seen to provide a good description of ATLAS data in the new phase space explored by the LHC during Run I.
Precision measurements of dijet production, and photon production with and without jets provide powerful constraints on the gluon PDF of the proton at high momentum fraction.
The measurement of the $\phi$(1020) meson complements these measurements by constraining the s-quark and gluon PDFs at low momentum fraction.
In addition, examining jet shapes in top-quark pair events provides an enriched sample of $b$-jet and light-jet parton showers.
This, along with further tuning of the UE, will be important for further refining the MC generators before beginning Run II at the LHC.
Dijet production has also been used to confront new models of physics, showing the importance of these measurements beyond refining QCD modeling.
Going forward, the measurements from 2012 will also be crucial for getting a strong start when data collection resumes in 2015.

\end{document}